\documentclass[a4paper]{article}
\usepackage[latin1]{inputenc} % ou \usepackage[utf8]{inputenc}
\usepackage{RR,RRthemes}
\usepackage{hyperref}

\usepackage{amssymb}
\usepackage{amsmath}
\usepackage{amssymb}
\usepackage{tabularx}
\usepackage{color}
\usepackage{epsfig}
\usepackage{wrapfig}
\usepackage{listings}
\usepackage{fancyvrb}
\usepackage{algorithmic}%
\usepackage{algorithm}

\usepackage[algo2e,ruled,lined]{algorithm2e}

\newtheorem{Define}{Definition}

\usepackage{graphicx}
\definecolor{gris25}{gray}{0.75}

\RRdate{May 2010}
\RRauthor{Nadjib Lazaar
\thanks[sfn]{INRIA Rennes Bretagne Atlantique, Campus Beaulieu, 35042 Rennes, France mail: \{lazaar.nadjib, arnaud.gotlieb\}@irisa.fr}
\and 
Arnaud Gotlieb
\thanksref{sfn}
\and
Yahia Lebbah
\thanks{Universit\'e d'Oran Es-Senia, B.P. 1524 EL-M'Naouar, 31000 Oran, Algerie Universit\'e de Nice--Sophia Antipolis, I3S-CNRS, France, mail: ylebbah@gmail.com}
}

\RRetitle{On Testing Constraint Programs}
\RRtitle{Vers le test des programmes à contraintes} 
\RRabstract{
The success of several constraint-based modeling languages such as OPL, ZINC, or COMET, appeals for better software
engineering practices, particularly in the testing phase. This paper introduces a testing framework enabling automated
test case generation for constraint programming. We propose a general framework of constraint program development
which supposes that a first declarative and simple constraint model is available from the problem  specifications
analysis. Then, this model is refined using classical techniques such as constraint reformulation, surrogate
and global constraint addition, or symmetry-breaking to form an improved constraint model that
must be thoroughly tested before being used to address real-sized problems. We think that most of the faults
are introduced in this refinement step and propose a process which takes the
first declarative model as an oracle for detecting non-conformities. We derive
practical test purposes from this process to generate automatically test data that exhibit non-conformities.
We implemented this approach in a new tool called CPTEST that was used to automatically detect
non-conformities on two classical benchmark programs, namely the Golomb rulers and the car-sequencing problem.}

\RRresume{
Tout processus de développement logiciel effectué dans un cadre industriel inclut désormais une phase de test ou  de vérification formelle, y compris pour le développement des programmes à contraintes. Notre travail vise à définir une plateforme de test qui génère automatiquement des données de test pour les programmes à contraintes. Cette nouvelle plateforme est également motivée par le développement récent de plusieurs langages de modélisation de haut-niveau tels que OPL, ZINC, COMET, ou GECODE, qui ouvre la voie à des recherches orientées vers les aspects génie logiciel autour de la programmation par contraintes. Notre approche repose sur certaines hypothèses quant au développement et raffinement dans un langage de PPC. Il est usuel de démarrer à partir d'un modèle simple et très déclaratif, une traduction fidèle de la spécification du problème, sans accorder d'intérêt à ses performances.
Par la suite, ce modèle est raffiné par l'introduction de contraintes redondantes ou reformulées, l'utilisation de structures de données optimisées, et de contraintes globales.
Nous pensons que l'essentiel des fautes introduites est compris dans cette deuxième étape.
Dans ce rapport nous bâtissons une plateforme de test des programmes à contraintes qui établit les règles de conformité entre le modèle initial déclaratif et le programme optimisé dédié à la résolution d'instances de grande taille.
Nous avons implémenté cette approche en un premier prototype de test appelé CPTEST. Cet outil permet  de détecter automatiquement des non-conformités entre le modèle de départ et le programme raffiné. Nous présentant également à la fin une première validation expérimentale sur deux problèmes connus, les règles de Golomb et le problème d'ordonnancement des véhicules (POV).

}
\RRkeyword{software testing, constraint programming, conformity relation, constraint negation}
\RRmotcle{test logiciel, programmation par contraintes, relation de conformité, négation de contrainte}
\RRprojets{Celtique}
\RRdomaine{2}
\URRennes
\RCRennes % Rennes - Bretagne Atlantique

\begin{document}
\RRNo{7291}

\makeRR   

\section{Introduction}

Constraint programs such as those written in modern Constraint Programming languages and platforms
(e.g. OPL\footnote{www.ilog.com/products/oplstudio/},  COMET\footnote{www.dynadec.com/support/downloads/},
ZINC \footnote{http://www.g12.cs.mu.oz.au/}, CHOCO\footnote{choco.sourceforge.net}, GECODE\footnote{www.gecode.org}, ...),
aim at solving industrial combinatorial problems that arise in optimization, planning, or scheduling. Recently,
a new trend has emerged that propose also to use CP programs to address critical applications in e-Commerce \cite{HS05},
air-traffic control and management \cite{FPA07,JV08}, and critical software development \cite{CRH08,Got}.  While constraint program debugging drew the attention of some researchers, few supports in terms of software engineering and
testing have been proposed to help verify critical constraint programs.
Automatic debugging of constraints programs has been an important topic of the
OADymPPaC\footnote{http://contraintes.inria.fr/OADymPPaC/} project, that resulted
in the definition of generic trace models \cite{Deransart,langevine},
the development of post-mortem trace analyzers, such as Codeine for
Prolog, Morphine \cite{langevine} for Mercury, ILOG Gentra4CP,
or JPalm/JChoco.
These models and tools help understand constraint programs and contribute to their optimization and
correction, but they are not dedicated to systematic fault detection.
Indeed, functional fault detection requires the definition of a reference (called an oracle
in software testing) in order to check the conformity between an implementation and its reference.
Automatic fault detection also requires the definition of test purpose to decide when to stop testing.
Whereas conventional software development benefits from research advances in software verification
(including static analysis, model checking or automated test data generation), developers of constraint programs
are still confined to perform systematic verification by hand.
 
  Automatic constraint program testing cannot be easily handled by existing testing
approaches because of the two following reasons: firstly, constraint programs are intrinsically non-deterministic
as they represent sets of solutions and conventional definitions of conformity do not apply~;
secondly, the refinement process of constraint programs is specific to CP.  Indeed, developers usually start with an initial declarative
constraint model of the problem, which faithfully translates the problem specifications,
without granting interest to its performances. As this model cannot handle large-sized instances
of the problem, they exploit several refinement techniques to build an improved model. For example, usual
refinement techniques include the use of dedicated data structures, constraint reformulation, global constraints addition,
redundant and surrogate constraint addition, as well as constraints
which break symmetries (these constraints usually improve considerably the effectiveness
of the solving process). The refinement process, carried out by the developer, is an error-prone process
and we believe that most of the faults are introduced during this step.

In this article, we propose a testing framework for checking the correction of a constraint program implementation.
The oracle for the constraint program under test is an initial declarative model considered to be valid w.r.t. the user requirements.
Our framework is based on the definition of four distinct conformity relations to handle constraint satisfaction problems as well
as optimization problems. A practical consequence of these definitions is
the proposal of test purposes for evaluating the conformance of constraint programs.
Note that this paper does not address another essential topic of CP verification which is the correction
of solvers or optimizers. We propose an algorithm for checking the correction of the CP program under test that
solves a set of derived constraint problems able to exhibit non-conformities. 
We implemented our approach in a tool called CPTEST that seeks non-conformities in OPL programs.
For evaluating the proposed testing process, CPTEST was used to find non-conformities in various faulty OPL constraint programs
of the Golomb rulers and the car-sequencing problem. It was also used to assess the conformity for small instances of the problem.

The rest of the paper is organized as follows: Sec. \ref{sec-motivating} illustrates our
testing framework on a simple case in order to show a typical non-conformity case.
Sec. \ref{sec-testing} gives the definition of conformity relations required in the framework.
In Sec. \ref{sec-cp-testing}, the testing process we derive from these definitions is introduced and illustrated on
a simple example. Sec. \ref{sec-experimental} presents the CPTEST tool and details our experimental evaluation.
Finally, Sec. \ref{sec-conclusion} concludes the paper and draws some perspectives to this work.

\section{An illustrative example}
\label{sec-motivating}

Let us illustrate some of the refinement techniques on the classical problem of the Golomb rulers,
which has various applications in fields such as Radio communications or X-Ray crystallography.

A Golomb ruler \cite{001} is a set of $m$ marks $0 = x_1 < x_2 <... < x_m$ such as $m (m-1)/2$
distances $\{x_j - x_i |\ 1 \leq i < j \leq m\}$ are distinct. A ruler is of order $m$ if it contains $m$ marks,
and it is of length $x_m$. The goal is to find a ruler of order $m$ with minimal length ($minimize\ x_m$).
\begin{figure*}[htb]
\scriptsize
\begin{minipage}{12.5cm}
\begin{Verbatim}[frame=single]
   int m=...;                            int m=...;                                                       
   dvar int+ x[1..m];                    dvar int x[1..m] in 0..m*m;                             
   minimize x[m];                        tuple indexerTuple { int i; int j;}
   subject to {                          {indexerTuple} indexes={<i,j>|i,j in 1..m: i < j};           
    c1: forall (i in 1..m-1)             dvar int d[indexes];
         x[i] < x[i+1];                  minimize x[m];
    c2: forall (i,j,k,l in 1..m :        subject to {
        (i < j && k < l &&             cc1: forall (i in 1..m-1) 
        (i != k || j != l)))                 x[i] < x[i+1];
        x[j] - x[i] != x[l] - x[k];    cc2: forall(ind in indexes)       
   }                                      d[ind] == x[ind.i]-x[ind.j]; 
                                       cc3: x[1]=0;
                                       cc4: x[m] >= (m * (m - 1)) / 2;
                                    // cc5: allDifferent(all(ind in indexes ) d[ind]);
                                       cc6: x[2] <= x[m]-x[m-1];
                                       cc7: forall(ind1 in indexes, ind2 in indexes, 
                                          ind3 in indexes: (ind1.i==ind2.i)&&
                                          (ind2.j==ind3.j) &&(ind1.j==ind3.j)&&
                                          (ind1.i<ind2.j < ind1.j)) d[ind1]==d[ind2]+d[ind3];
                                       cc8: forall(ind1,ind2,ind3,ind4 in indexes: 
                                          (ind1.i==ind2.i)&&(ind1.j==ind3.j)&&
                                          (ind2.j==ind4.j)&&(ind3.i==ind4.i)&&(ind1.i<m-1)
                                          &&(3<ind1.j<m+1)&&(2<ind2.j<m)&&(1<ind3.i<m-1)&&
                                            (ind1.i < ind3.i < ind2.j < ind1.j))
                                          d[ind1]==d[ind2]+d[ind3]-d[ind4];
                                       cc9: forall(i in 2..m, j in 2..m, k in 1..m : i  < j)
                                          x[i]=x[i-1]+k => x[j] != x[j-1]+k;
                                        }
            - A -                                         - B -
\end{Verbatim}
\end{minipage}
%\hspace{1cm} \textbf{- A -}\hspace{6cm}  \textbf{- B -}
\caption{\footnotesize $M_x(k)$ and $P_x(k)$  of  Golomb rulers problem in OPL.}
\label{v0}
\end{figure*}
A declarative model of this problem is given in part {\tt A} of Fig.\ref{v0} while
part {\tt B} presents a refined and improved model. It is easy to convince a human that model {\tt A} actually
solves the Golomb rulers problem, but this is much more difficult for model {\tt B}.
Indeed, model {\tt B} uses a matrix as data structure ({\tt d[indexes]}), statically breaks symmetries ({\tt cc6}), it contains
redundant and surrogate constraints ({\tt cc7,cc8,cc9}) and global constraints ({\tt allDifferent}).
In this paper, we address the fundamental question of revealing non-conformities in between the constraint
program under test {\tt B} and the model-oracle {\tt A}. Testing {\tt B} before using it on large instances of
the problem (when $m>15$) is highly desirable as computing the global
minimum of the problem for these instances may require computation time greater than a week. Note that {\tt B} is syntactically
correct and provides correct Golomb rulers for small values of $m$.
Our testing framework tries to find an instantiation of the variables that satisfies the
constraints of {\tt B} and violates at least one constraint of {\tt A}. This testing process is detailed in section \ref{sec-cp-testing}.
With $m=8$, our CPTEST framework computes $x =  [0\ 1\ 3\ 6\ 10\ 26\ 27\ 28]$ in less than $6$sec on a standard machine,
indicating that {\tt B} does not conform {\tt A} and then contains a fault.
Indeed, $x$ is not a Golomb ruler as $27-26 = 1- 0 =1$. In fact, this non-conformity
can easily be tackled by removing the comment on constraint {\tt cc5} in part {\tt B}. Doing so;
CPTEST provides a conformity certificate saying that the CP program actually computes the global minimum 
in $10034.69$sec (about 3hours). However, note that this certificate is only valid for $m=8$. Note also that our framework
can handle non-conformities of the Golomb rulers where the global minimum requirement is relaxed in order
to deal with larger instances (when $m>30$).
\section{Testing constraint programs}
\label{sec-testing}

\subsection{Notations}
In the rest of the paper, $x$ denotes a vector of variables and $(x\backslash x_i)$ stands for substituting $x$ by the valuation 
$x_i$. 

\begin{wrapfigure}{l}{20mm}
  \centering
\footnotesize
\underline{Model $M_x(k)$}\\
$\begin{cases} 
C_1(x)\\
...\\
C_n(x)\\
\end{cases}$\\
$Solve()$\\
\normalsize
\end{wrapfigure}
\begin{wrapfigure}{r}{41mm}
  \centering
    \leavevmode \epsfxsize=4cm\mbox{{\epsfbox{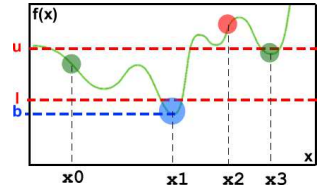}}}
\caption{\footnotesize Objective solutions.\normalsize}
\label{f}
\end{wrapfigure}

A constraint program includes a constraint model $M_x(k)$, which is a conjunction of constraints 
$C_i(x)$ over variables $x$ parameterized by $k$, the parameters vector of the model. Note that
$x$ may depend on $k$.
For the Golomb rulers, $k$ is the order of the ruler while $x$ represents the vector of marks.
If $k=3$ then one seeks for a ruler with $3$ marks (e.g., {\tt x=[0 1 3]}) while if $k=4$ one 
seeks for a ruler with 4 marks (e.g., {\tt x=[0 1 4 6]}).
$Solve()$ is a generic procedure representing either the call to a constraint solver in the case of constraint satisfaction
problem or the call to an optimization procedure. In this latter case, we note $f$ the cost function
(for the sake of clarity, $f$ will be a minimization function but maximization problems can be tackled as well).
We consider that $k$  belongs to $\mathcal{K}$ the set of possible values of the parameters for which 
$M_x(k)$ has at least one solution. $sol(M_x(k))$ denotes the set of solutions of $M_x(k)$ and while 
$Proj_y(sol(M_x(k)))$ expresses the projection of $sol(M_x(k))$ on the set $y$ when $y\subseteq x$. 
In optimization problems, one usually starts with feasible solutions ranging in a cost interval $[l, u]$. Therefore, we introduce the set
\[Bounds_{f,l,u}(M_x(k)) = \{ x | x \in sol(M_x(k)), f(x) \in [l,u]\}\]
To clarify these notations, Fig. \ref{f} shows an example of a real objective function where point $x_1$ is a global minimum 
with a cost $f(x_1)=b$ and points $x_0, x_3$ belongs to $Bounds_{f,l,u}(M_x(k))$. Note that $x_1$ as well as $x_2$ do not necessarily belong to $Bounds_{f,l,u}(M_x(k))$.

\subsection{Constraint models and programs}

In our framework, we consider the initial declarative constraint model to be a testing oracle, called the \textsl{Model -- Oracle }, 
and noted $M_x(k)$. $M_x(k)$ represents all the solutions of the problem and strictly conforms the problem specifications. 
We suppose that, for any parameter instantiation, $M_x(k)$ possesses at least one solution.
Considering unsatisfiable Model--Oracles could be interesting for some applications (such as software verification \cite{Got}) 
but we excluded these cases in order to avoid considering equivalence of unsatisfiable models. 
The \textsl{Constraint Program Under Test (CPUT)} is a constraint model $P_z(k)$ (possibly unsatisfiable) which has to 
be tested for correction against the Model--Oracle. $P_z(k)$ is intented to solve difficult instances of the problem.
%it is highly desirable to test it before using it on these instances, as difficult instances usually require large resources. 
We built our framework on the hypothesis that checking whether $M_{(x\backslash x_0)}(k_0)$ is true where $x_0$ 
is a point of the search space is not hard, while finding such an $x_0$ satisfying the constraints may be hard.
Given a CPUT $P_z(k)$ and its Model-Oracle $M_x(k)$, we suppose that $x \subseteq z$ as $P_z(k)$ was obtained by refining $M_x(k)$.
Hence, the set of variables in $z$ distinct of $x$ are dependant variables that are automatically instantiated when $x$ is instantiated.

\subsection{Conformity relations}
The correction of a CPUT w.r.t. a Model--Oracle can be approached through the usage of conformity relations. 
These relations aim at assessing the correction of the CPUT, a notion that can be expressed with various levels of 
depth. We propose four set-based definition of conformity divided on two groups: conformity relations adapted to
constraint satisfaction problems and conformity relations for optimization problems.

\subsubsection{Conformity relations for constraint satisfaction problems}\label{cs}
\begin{figure}[htbp]
  \begin{center}
   \includegraphics[scale=0.20]{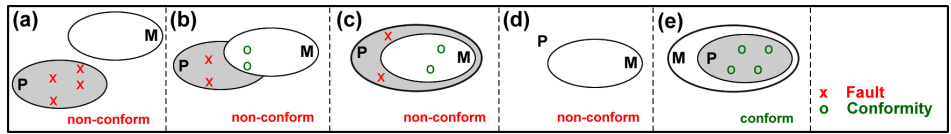}
    \caption{\footnotesize $conf_{one}$ on $P_z(k)$ and $M_x(k)$.\normalsize}
    \label{one}
  \end{center}
\end{figure}

The simplest definition of correction, well-adapted for problems where a single solution is sought, is given by the following 
conformity relation:
\begin{Define}[$conf_{one}$]  
\begin{equation*}
\begin{array}{l}
P\ conf^k_{one}\ M\ \Leftrightarrow  Proj_x(sol(P_z(k)))\neq \emptyset\ \wedge\ Proj_x(sol(P_z(k))) \subseteq sol(M_x(k)) \\
P\ conf_{one}\ M\  \Leftrightarrow\ (\forall k\in \mathcal{K}, P\ conf^k_{one}\ M)
\end{array}
\end{equation*}
\end{Define}
Roughly speaking, for a given instance $k$, $conf^k_{one}$ asks the solutions of the CPUT to be included in the solutions of the Model-Oracle.
As an example, Fig.\ref{one} presents both the sets $sol(M_x(k))$ noted M and $sol_x(P_z(k))$ noted P, where 
points in red \textbf{x} raise non-conformities (i.e., faults in the CPUT) while points in green \textbf{o} are conform w.r.t. the Model--Oracle.
Parts (a)(b)(c) of Fig.\ref{one} exhibit non-conformities as solving $P_z(k)$ can lead to solutions which do not satisfy $M_x(k)$. 
Part (d) does not exhibit any non-conformity but, as P does not contain any solution, it does not conform the Model--Oracle for $conf_{one}$. 
This example also shows that unsatisfiable models must be considered as non-conform w.r.t. Model--Oracles, in order to tackle faulty 
unsatisfiable CPUTs.
On the contrary, part (e) of Fig.\ref{one} shows that $P_z(k)$ conforms $M_x(k)$ for $conf_{one}$, as P cannot contain any non-conformity points. 

\vspace{-0.5cm}

Whenever all the solutions are sought, another definition of conformity is useful:
\begin{Define}[$conf_{all}$]  
\begin{equation*}
\begin{array}{l}
P\ conf^k_{all}\ M\ \Leftrightarrow  Proj_x(sol(P_z(k))) = sol(M_x(k))\ (\neq \emptyset) \\
P\ conf_{all}\ M\  \Leftrightarrow\ (\forall k\in \mathcal{K}, P\ conf^k_{all}\ M)
\end{array}
\end{equation*}
\end{Define}
Roughly speaking, $conf_{all}$ asks for both set of solutions to be the same. Satisfying this conformity relation is very
demanding and not always pertinent. For instance, the CPUT in part {\tt B} of Fig.\ref{v0} includes constraints that break
symmetries of the problem (e.g., {\tt cc6}), which yields to lose solutions from the Model-Oracle. As a result, those 
two models cannot be conform w.r.t. $conf_{all}$.   
\begin{figure}[htbp]
  \begin{center}
   \includegraphics[scale=0.20]{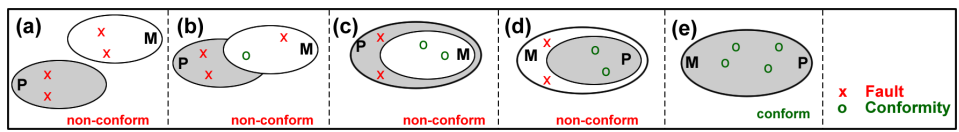}
			\caption{\footnotesize $conf_{all}$ on $P_z(k)$ and $M_x(k)$.\normalsize}
    \label{all}
  \end{center}
\end{figure}
In Fig. \ref{all}, parts (a)(b)(c) and (d) exhibit non-conformities. Part (d) shows a solution 
of the Model--Oracle which is not solution of the CPUT~; therefore, the CPUT is a faulty over-constrained model. 
Part (c) exhibits the opposite case where the CPUT is a faulty under-constrained model.
Proving that $P_z(k)$ conforms $M_x(k)$ for one of these two conformity relations is highly desirable.
Unfortunately, such a proof would require not only to find all the solutions of the CPUT which
is an NP\_hard problem for some constraint languages (e.g., the finite domains constraint language), but also
to perform this for any value of $k$. This seems to be intractable in general (probably undecidable) and
then we will confine ourselves to the search of non-conformities within finite resources. 

\subsubsection{Conformity relations for optimization problems}\label{optim}
Conformity relations for optimization problems is harder to define, as practicians usually start their
refinement process by the definition of bounds for the optimal case \cite{ST02} . Note also that non-conformities
may arise in the cost function itself and we wanted our conformity relations to be able to tackle those cases. 
\begin{figure}[htbp]
  \begin{center}
   \includegraphics[scale=0.20]{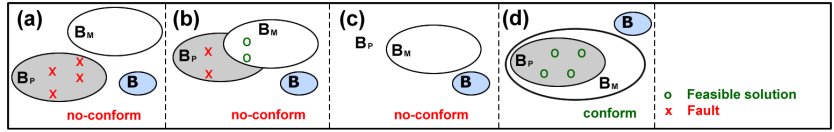}
    \caption{\footnotesize $conf_{bounds}$ on $P_x(k)$ and $M_x(k)$.\normalsize}\label{bounds}
    \label{bounds}
  \end{center}
\end{figure}
\vspace{-0.5cm}

Fig.\ref{bounds} presents the conformity relation where feasible solutions of the CPUT are sought in $[l,u]$. 
$B_P$ denotes the set $Bounds_{f,l,u}(P_x(k))$, $B_M$ denotes the set $Bounds_{f,l,u}(M_x(k))$ while 
B is the set of global minima of $M_x(k)$.
Part (a) exhibits four non-conformities as these points are not feasible solutions of the Model--Oracle $M_x(k)$ in $[l,u]$.
For the same reason, Part (b) exhibits two non-conformities as two feasible solutions of $B_P$ with cost in $[l,u]$ do not belong to $B_M$.  
Part (c) presents also a non-conformity as $B_P$ does not contain any feasible point meaning that the minimization problem
cannot find a feasible solution with cost in $[l,u]$. 
On the contrary, part (d) shows conformity because solutions of $B_P$ belong to $B_M$. 
Formaly speaking,
\begin{Define}[$conf_{bounds}$]
\begin{equation*}
\begin{array}{l}
P\ conf^k_{bounds}\ M\ \Leftrightarrow Proj_x(bounds_{f,l,u}(P_z(k)))\neq \emptyset\ \\
\qquad \qquad \qquad \qquad \quad \wedge\ Proj_x(bounds_{f',l,u}(P_z(k))) \subseteq bounds_{f,l,u}(M_x(k)) \\
\end{array}
\end{equation*}
\end{Define}
Note that the definition of $conf_{bounds}$ does not require that $f=f'$ and then cases where the cost function has been refined can
also be handled. This conformity relation is useful for addressing hard optimization problems as it does not require the computation
of global minima. As a result, it can be used to assess the correction of models on relaxed instances of the global 
optimization problems. We will come back on this advantage in the experimental validation section.
However, for some problems, it may be useful to assess not only the correction but also the fact that the CPUT actually computes
optimal solutions. This can be performed by using the following definition which ensures that the global optimum belongs to $[l, u]$.
\begin{Define}[$conf_{best}$]
\begin{equation*}
\begin{array}{l}
P\ conf^k_{best}\ M\  \Leftrightarrow\ \begin{cases} P\ conf^k_{bounds}\ M, \\
bounds_{f,-\infty,l}(M_x(k))=\emptyset, \\
bounds_{f',-\infty,l}(P_z(k))=\emptyset\\
\end{cases} 
\end{array}
\end{equation*}
\end{Define}

\section{A CP testing framework}
\label{sec-cp-testing}

Testing a CPUT w.r.t. an model-oracle requires to select test data. In this context,
a test datum defines an instance of the CPUT and a point of the search space.
\begin{Define}[Test datum]
Given a CPUT $P_z(k)$ and a Model--Oracle $M_x(k)$, a {\it test datum} is an instantiated pair $(k_0,x_0)$ of parameters and variables.
\end{Define}
Note that evaluating $M_k(x)$ on the test datum $(k_0,x_0)$ results true when $x_0$ is a solution of the model and false otherwise.
Test execution is realized by evaluating both $P_{z \backslash z_0}(k_0)$ and
$M_{x \backslash x_0}(k_0)$\footnote{$z_0$ is obtained by extending $x_0$ with values depending on $x_0$}
and checks whether the results (either true or false) are the same. Depending on the selected conformity relation, a test verdict can be issued.
This elementary process can be repeated as long as one wishes, but it is more interesting to guide the
test data generation process by the use of {\it test purposes}.
%\subsection{Criteria based on constraint negation}
Seeking non-conformities implies finding test data such as the CPUT is satisfied and the Model--Oracle is violated.
This enables to detect faults in CPUT, and helps the constraint programmer to revisit its refinements.
Based on the selection of a conformity relation, non-conformities can be sought with the following test purposes:
\begin{description}
\item[$conf_{one}$] Given $k$, find a solution to $P_z(k)\wedge \neg C_i$ where $C_i$ is a constraint
of the Model-Oracle $M_x(k)$. The idea here is to isolate a non-conformity by looking independently
at each constraint of the model-oracle. Considering all the constraints of the model-oracle would also be
possible but less efficient to detect non-conformities as more constraints would be involved. Note that
heuristics can be defined on the order of constraints to consider first. Note also that proving the unsatisfiability of
$P_z(k)\wedge \neg C_i$ for all $C_i \in M_x(k)$ permits to issue a {\it conformity certificate} saying that $P conf^k_{one} M$.
\item[$conf_{all}$] Given $k$, find a solution to
$(M_x(k)\wedge \neg C'_i) \vee (P_z(k)\wedge \neg C_i)$ where $C_i$ (resp. $C'_i$) is a constraint
of the Model-Oracle $M_x(k)$ (resp. $P_z(k)$). In this case, proving the unsatisfiability of these constraints
permits to issue the conformity certificate $P conf^k_{all} M$, but this is not often desirable as constraint solving usually requires to
issue a single solution instead of all solutions.
\item[$conf_{bounds}$] Given $k$ and $[l,u]$, find a solution to  $P_z(k)\wedge \neg C_i \wedge f'(z)\in[l,u] \wedge f(x)\in [l,u]$ where
$f, f'$ are the cost functions
of the Model-Oracle $M_x(k)$ and the CPUT $P_z(k)$. Proving that these constraints are unsatisfiable permits to issue a
certificate $P conf^k_{bounds} M$.
\item[$conf_{best}$] Given $k$, find a solution to
$(P \neg conf^k_{bounds} M) \vee bounds_{f,-\infty,l}(M_x(k))\neq\emptyset \vee bounds_{f,-\infty,l}(P_z(k))\neq\emptyset$.
Proving that these constraints are unsatisfiable permits to issue a conformity certificate $P conf^k_{best}$.
\end{description}
Interestingly, any solution found by the guidance of one of these test purposes can be stored for further investigations.
Indeed, it can be used to debug the CPUT by looking at the violated constraint and it can also enrich a test set that
will serve to assess the correction of future versions of the CPUT. In addition, conformity certificates are essential for
those who want to convince third-party certification authorities that their CP programs can be used in critical systems \cite{HS05,Got}.
So, the proposed testing framework has a role to play
in various phases of the constraint program development.

We now propose a simple but generic algorithm for searching non-conformities
(Algorithm \ref{algo-nonconf}).

\scriptsize

\renewcommand{\algorithmicrequire}{\textbf{Input:}}

\renewcommand{\algorithmicensure}{\textbf{Output:}}

\begin{algorithm2e}[H]

%\SetLine

\SetKwFunction{Divide}{Divide}

\SetKwInOut{Input}{Input}

\SetKwInOut{Output}{Output}

\caption{{one\_negated($D$, $\{C_1,...C_n\}$)}}\label{algo-nonconf}

\Input{$D$, $\{C_1,...C_n\}$ set of constraints.}

\Output{$conf$ when $\{C_1,...C_n\}$ conform $D$, $\neg conf$(+ non-conformity\ point) otherwise}

\BlankLine

$nc\leftarrow \emptyset$ \qquad

$X\leftarrow vars(D)$\\

\ForEach{ $C_i \in \{C_1,...,C_n\}$ }{

$V\leftarrow vars(C_i)/X$\\

\lIf{$V =\emptyset$}{
$nc \leftarrow Solve( D \wedge \neg C_i)$\\
}

\lElse{
$nc \leftarrow Solve( D \wedge \neg Proj_X(C_i))$\\
}

\lIf{$nc$}{\Return {$\neg conf(nc)$}}

}

\Return{ $conf$}

\end{algorithm2e}

where $Solve(D)$ denotes the algorithm to find the first solution of the constraints $D$, $vars(D)$ denotes the set of variables in $D$ and $Proj_X(C)$ denotes the constraint projection on variables $X$.\\

\normalsize
Algorithm \ref{algo-nonconf} takes two constraint sets as input and returns either $conf$ when both sets conform with relation $conf_{one}$
or $\neg conf$(non-conformity\ point) where a non-conformity point has been found. Note that the other conformity relations
can easily be implemented using this algorithm just by adjusting the call parameters.
Special care has to be taken when building the negation of a model.
For example, consider a Model-Oracle M with {\tt x-y!=x-z;   x-y!=y-z;  x-z!=y-z;} and a CPUT P with {\tt c1: x-y=d1;  c2: x-z=d2; c3: y-z=d3;
c4: allDiff(d1,d2,d3);}.
Here, it is trivial to see that  $P conf_{all} M$ but if {\tt c1} is
selected for negation, $M \wedge \neg c1$ has solutions as
{\tt d1} is out of the scope of M. In the definitions of the
conformity relations, these cases were discarded by the use
of projections on the variables of the model-oracle. As computing general projections
is expensive, pragmatic solutions are available in our implementation (see Sec.\ref{sec-experimental}).

Algorithm \ref{algo-nonconf} is the core algorithm of the presented testing framework and
several implementation improvements are described in Sec.\ref{implem}. Providing that the underlying
constraint solver is sound and complete, this algorithm is sound as it cannot report $conf$ if there exists a non-conformity point. 
Indeed, given $k$, upon completion of the algorithm the unsatisfiability of $P_z(k) \wedge \neg M_x(k)$ is demonstrated showing that both models
conform with the selection conformity relation. It is also complete as it cannot report
false non-conformities. 

A keypoint of our approach is that test data can be automatically generated using the same constraint solver as the one used
for solving the CPUT. Recall that we rely on the solver and we are only interested in detecting non-conformities in models.

\section{Experimental validation}
\label{sec-experimental}

\subsection{Implementation}\label{implem}
We implemented the testing framework shown above in a tool called CPTEST for OPL (Optimization Programming Language \cite{Hen99}).
We chose OPL because it is one of the main programming environments for developing constraint programs and
also critical constraint programs \cite{FPA07}. CPTEST is based on ILOG CP Optimizer 2.1 from ILOG
OPL 6.1.1 Development Studio. All our experiments were performed on Quadcore IntelXeon 3.16Ghz machine with
16GB of RAM and all the models we used to perform these experiments are available online at {\url www.irisa.fr/celtique/lazaar/CPTEST}.

CPTEST includes a complete OPL parser and a backend process that produces dedicated OPL programs as output. 
These OPL programs must be solved
in order to find non-conformities. If a solution is found, then CPTEST stops and reports the non-conformity to the user.
Whenever all these OPL programs are shown to be inconsistent, then
a conformity certificate is issued. The tool is parameterized by several options, including the chosen conformity relation,
the instance of the problem, etc. CPTEST handles the overall OPL language
and can negate most of the constraints that can be expressed in OPL. However, it cannot
negate all the global constraints available, such as the {\tt cumulative} or {\tt circuit} global constraint.
Tab.\ref{synt} summarizes the syntax of OPL constraints handled by CPTEST.
\begin{table}[t]
\caption{Syntax of OPL expressions handled by CPTEST}\label{synt}
\scriptsize
\begin{tabular}{ll}
\hline
\\
$Ctrs$ {\tt ::=} & $\ \ \ Ctr \ | \ Ctrs$ \\
$Ctr$  {\tt ::=} & $\ \ \ rel \ | \ $ {\tt forall(} $rel$ {\tt )} $Ctrs  | \ $ {\tt or(} $rel$ {\tt )} $Ctrs \ | \ $ {\tt if(} $rel$ {\tt )} $Ctrs$ {\tt else} $Ctrs$\\
                 & $| \ $ {\tt allDifferent(}$rel${\tt )} $| \ $ {\tt allMinDistance(}$rel${\tt )} $| \ ${\tt inverse(}$rel${\tt )}$| \ $ {\tt forbiddenAssignments(}$rel${\tt )}\\
                 &$| \ $ {\tt allowedAssignments(}$rel${\tt )}$| \ $ {\tt pack(}$rel${\tt )}\\
\\
\hline
\end{tabular}
\end{table} 
OPL includes two aggregators, namely {\tt forall} and {\tt or}.
The universal qualifier {\tt forall} is used to declare a collection of closely 
related constraints and to build global constraints. Interestingly, the {\tt or} aggregator can be used 
to negate {\tt forall}, as {\tt or} implements existencial quantification.
The OPL {\tt If-then-else} statement is less general than it may appear as its condition cannot contain decision variables. 
Its negation can be computed by negating the Then-part and Else-part without any loss of generality, as our goal is only to find
non-conformities instead of computing the negation of a general model. 
Our CPTEST tool handles several global constraints over discrete values, namely 
{\tt \footnotesize allDifferent, allMinDistance, inverse, forbiddenAssignments, allowedAssignments} 
and {\tt \footnotesize pack}. These constraints can be represented as an aggregation of constraints and then computing their
negation becomes trivial with the rules presented above and using the other global constraints.
For example, the negation of {\tt C: allDifferent(all(i in R) x[i]) } is {\tt or(ordered i,j in R) x[i] = x[j]}
as {\tt C} rewrites to {\tt forall(ordered i,j in R) x[i] != x[j]}, and the negation of
{\tt \footnotesize forbiddenAssignments} is simply {\tt \footnotesize allowedAssignments}.

We implemented algorithm \ref{algo-nonconf} in CPTEST with several improvements. In particular, by noticing that it is unnecessary
to search for non-conformities on constraints that are included in both the CPUT and the Model-Oracle, we implemented
a simple rewriting system to check equality modulo Associativity-Commutativity ($\equiv_{AC}$). The system implements
the following rules:\\
$\begin{Bmatrix}
x\circ y\rightarrow y\circ x,& (x\circ y)\circ z\rightarrow x\circ (y\circ z),& x+0\rightarrow x,\\
x*1\rightarrow x,& x*0\rightarrow 0,&  x \times (y\bullet z)\rightarrow (x\times y)\bullet(x\times z),\\
x<y\leftrightarrow y>x,& x\leq y\leftrightarrow y\geq x,& x-0\rightarrow x,\\
\end{Bmatrix}$\\
where $\circ \in \{+,*,\wedge,\vee\}$, $\times \in \{*,\wedge,\vee\}$ and $\bullet \in \{+,\wedge,\vee\}$.

In algorithm \ref{algo-nonconf}, the constraint $C_i$ is discarded whenever there exists $C'i$ in $D$ such as  
$C'_i \equiv_{AC}(C_i)$. \\
We have seen in sec.\ref{sec-cp-testing} that computing general projection is expensive, 
we can enumerate some practical solutions to handle local variables and the constraint projection:
\begin{itemize}
	\item Annotating constraints of CPUT.
	\item computing projections (Fourier elimination).
	\item checking non-conformities.
\end{itemize}

It is important to stress that projections are computing when we seek all solutions ($conf_{all}$)  
and we have $C_i\in P$ to negate ($M \wedge \neg C_i$).
We implement in CPTEST the first and the last proposed solution. The CPTEST user's can annotate his 
CPUT by indicating the constraint that connects base and local variables. Otherwise, CPTEST check 
if the non-conformity reached is a real one or a false alarm. \\
The goal of our experimental evaluation was to check that CPTEST is able to detect faults in OPL programs.
We feeded CPTEST with faulty models coming from initial constraint program development.
Indeed, we developed optimized models of two well-known CP problems, namely the Golomb rulers
and the car sequencing problem, and we kept first versions of these models for which faults were found. 
\subsection{The Golomb ruler problem}
\label{sec-golomb}
The model-oracle of the Golomb rulers is given in part {\tt A} of Fig.\ref{v0} while part {\tt B} contains
a conform version of an optimized version of the model when the comment on constraint {\tt cc5} is removed.
Let us call {\tt P} this version.
The four intermediate versions of the Golomb rulers we kept from our initial program development contain
realistic faults, not invented for the experiment.
Tab.\ref{faults} shows the four faulty versions expressed with the constraints of {\tt P}.
Note that constraint {\tt cc6} breaks symmetries in the problem
and then it removes solutions (valid Golomb rulers) w.r.t. the model-oracle.
\scriptsize
\begin{table}[t]
\begin{center}
\caption{Faulty versions of the Golomb Ruler}
\begin{tabular}{|c|l|}\hline
& constraints of P present in the CPUT\\ \hline
\textbf{CPUT1}& {\tt cc1, cc9}\\
\textbf{CPUT2}& {\tt cc1, cc2, cc7, cc9}\\
\textbf{CPUT3}& {\tt cc1, cc2, cc7, cc8, cc9}\\
\textbf{CPUT4}& {\tt cc1, cc2, cc3, cc4, cc6, cc7, cc8, cc9, cc10}\\ \hline
\end{tabular}\label{faults}
\end{center}
\end{table}
\normalsize
Constraint {\tt cc10} is not documented in {\tt P}, it corresponds to {\tt forall( i in m..3*m) count(all(j in indexes)d[j],i)==1}. 
For each CPUT, we studied its conformity w.r.t. the model-oracle (part {\tt A}) using the four
conformity relations. The results we got for an instance parameter $m=8$ are given in Tab.\ref{golomb}.
For the $conf_{bounds}$ relation, the interval $[50, 100]$ was used to feed the relation, knowing
that the global minimum is $x_m=34$ when $m=8$.
Each time a non-confirmity was found, it was reported
with the CPU time required to find it. Firstly, the four faulty CPUT were reported as
being non-conforms and the time required for finding these
non-conformities is acceptable (less than a few minutes in the worst case). Secondly, this experiment
shows that the most practical conformance relations (i.e., $conf_{one}$ and $conf_{bounds}$) are preferable
to the other ones for efficiency reason. Indeed, for the first three CPUT, these relations gave results less than
$10$sec. Note that non-conformities are represented either by invalid Golomb rulers (e.g., $44-35 = 35 - 26 = 9$
in the CPUT1/$conf_{one}$ case) or by valid Golomb rulers (e.g., CPUT1/$conf_{all}$ case). In fact, a valid Golomb ruler $r$
can be produced when the model-oracle is satisfied by $r$ while the CPUT is refuted by $r$. These non-conformities
correspond to cases where the CPUT misses solutions of the problem. 
\begin{table}[t]
\caption{Non-conformities found by CPTEST in various CPUTs of the Golomb rulers problem (timeout = 1h30).}
\begin{tabular}{|c|c||c|c||c|c|}
\hline
\multicolumn{2}{|c||}{m = 8 }& $conf_{one}$ & $conf_{all}$& $conf_{bounds}$ & $conf_{best}$\\ \hline \hline
\multicolumn{2}{|c||}{ \tiny Non-conf points} & \tiny[0 7 8 18 24 26 35 44] & \tiny[17 18 20 25 34 45 49 55] & \tiny[0 2 3 6 11 58 72 86] & \tiny[0 1 3 6 10 15 24 33]\\ \cline{2-6}
{\textbf CPUT1} & \tiny T(s)& \tiny 4.29s & \tiny 21.45s & \tiny 5.64s & \tiny 7.31s\\ \hline \hline
\multicolumn{2}{|c||}{ \tiny Non-conf points}  & \tiny[0 4 5 26 28 31 47 63] & \tiny[17 18 20 25 34 45 49 55] & \tiny[0 18 39 43 45 46 55 64] & \tiny[0 3 4 9 13 15 24 33]\\ \cline{2-6}
{\textbf CPUT2} & \tiny T(s)& \tiny 5.62s & \tiny 40.78s & \tiny 4.64s & \tiny 174.43s\\ \hline \hline
\multicolumn{2}{|c||}{ \tiny Non-conf points}  & \tiny[0 4 5 26 28 31 47 63] & \tiny[0 4 5 26 28 31 47 63] & \tiny[0 18 39 43 45 46 55 64] & \tiny[0 3 4 9 13 15 24 33]\\ \cline{2-6}
{\textbf CPUT3} & \tiny T(s)& \tiny 9.53s & \tiny 45.78s & \tiny 7.15s & \tiny 389.04s\\ \hline \hline
\multicolumn{2}{|c||}{ \tiny Non-conf points}  & \tiny [0 12 18 20 29 33 34 39] & \tiny[1 2 10 22 33 55 57 60]& \tiny [0 21 30 32 42 45 46 50] & \tiny [0 6 13 21 22 25 27 32] \\ \cline{2-6}
{\textbf CPUT4} & \tiny T(s)& \tiny 12.60s & \tiny 0.15s & \tiny   9.01s & \tiny 12.53s \\ \hline
\multicolumn{2}{|c||}{ \tiny Non-conf points}  & \tiny conf & \tiny[0 7 9 12 37 54 58 64]& \tiny conf & \tiny --- \\ \cline{2-6}
{\textbf P} & \tiny T(s)& \tiny 3 448.46s & \tiny 0.18s & \tiny   3 658.13s & \tiny timeout \\ \hline
\end{tabular}
\label{golomb}
\end{table}
Interestingly, P is shown as being non-conform with the $conf_{All}$ relation and the non-conformity that is found
represent a valid Golomb ruler (i.e., [0 7 9 12 37 54 58 64]). In fact, recalling that P includes constraints that break the
symmetries, this result was expected. Finally, note that conformity of P when $conf_{best}$ is selected 
was impossible to assess within the allocated time (timeout=1h30). In fact, computing the global minimum of the Golomb ruler rapidly
becomes hard even for small values of $m$ (e.g., CPUT3/$conf_{best}$).  

\begin{figure}[htbp]
  \begin{center}
   \includegraphics[scale=0.30]{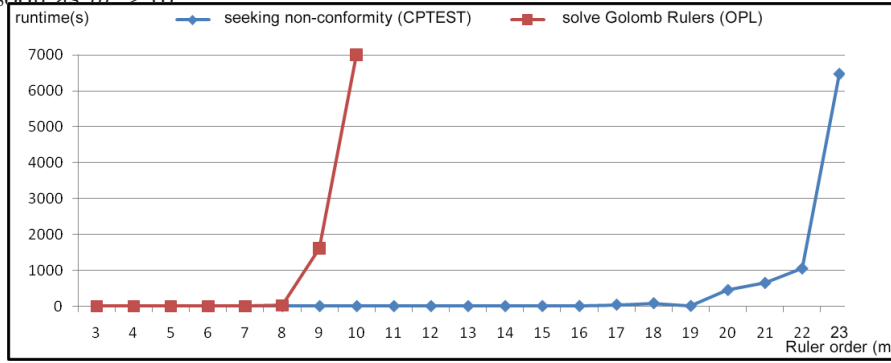}
    \caption{\footnotesize Testing time and solving time comparison on the Golomb rulers.\normalsize}
    \label{graphic}
  \end{center}
\end{figure}

Our experimental evaluation also had the goal to check that computing non-conformities with CPTEST was less hard than
computing solutions. For that, we compared the CPU time required to find non-conformities with $conf_{one}$ when the 
parameter value $m$ increases and the time required to solve Golomb on these instances.
Fig.\ref{graphic} shows that finding non-conformities with CPTEST remains tractable until $m=23$ while solving the CPUT
becomes intractable as soon as $m>10$. 

\vspace{-0.5cm}
\vspace{-1.5cm}

\subsection{The car sequencing problem}
\label{sec-car}

The car sequencing problem (CSeq) illustrates interesting features of CP including
wide parameter settings, redundant, surrogate and global constraints addition, and
specialized data structures definition. This is a constraint satisfaction problem that amounts 
to find an assignment 
of cars to the slots of a car-production company, which satisfies capacity constraints.

As a model-oracle of this problem, we took the model given in the OPL book \cite{Hen99}.
In this model, capacity constraints  are formalized by using constraints {\tt r outof s}, saying that from 
each sub-sequence of {\tt s} cars, a unit can produce at most {\tt r} cars with a given option.
Starting from this model, we built an optimized model by introducing several refinements, including
a new data structure {\tt \footnotesize setup[o,s]} which takes value 1 if option {\tt o} is installed on slot s,
redundant and global constraint addition (e.g., {\tt pack} constraint). When building our improved model
of car sequencing, we recorded four faulty constraint models that are used for experiments. Here again, the
idea was to keep models that represent realistic faults instead of a posteriori injected faults. 
These four models are available online on the site mentioned above.

\begin{table}[t]
\caption{Non-conformities found by CPTEST in various CPUTs of the car sequencing problem (timeout = 1h30).}\label{CSeq}
\begin{tabular}{|c |c ||c|c|| c|c|}
\hline
\multicolumn{2}{|c||}{ }&\multicolumn{2}{|c ||}{ \textbf {$Conf_{one}$} } & \multicolumn{2}{|c|}{ \textbf {$Conf_{all}$} }\\ \cline{3-6}
\multicolumn{2}{|c||}{  } & {\tt 10 slots} & {\tt 55 slots} & {\tt 10 slots} & {\tt 55 slots}\\ \hline
&{ \tiny Non-conf points} & \tiny 4 5 3 6 4 6 5 1 3 2 & \tiny\quad \qquad p1\qquad \quad\ \ \ \ \ & \tiny 4 5 4 6 3 6 5 1 3 2 & \tiny ---\\ \cline{2-6}
{\textbf CPUT1} & \tiny T(s)& \tiny 0.30s & \tiny 1.23s & \tiny 2.49s & \tiny timeout\\  \hline \hline
&{ \tiny Non-conf points}  & \tiny 4 6 3 1 5 2 3 5 4 6 & \tiny p2 & \tiny 5 4 3 5 4 6 2 6 3 1 & \tiny ---\\ \cline{2-6}
{\textbf CPUT2} & \tiny T(s)& \tiny 0.85s & \tiny 1.65s & \tiny 1.20s & \tiny timeout\\  \hline \hline
&{ \tiny Non-conf points}  & \tiny 5 2 3 6 1 4 3 6 4 5 & \tiny p3& \tiny  5 4 3 5 4 6 2 6 3 1  & \tiny --- \\ \cline{2-6}
{\textbf CPUT3} & \tiny T(s)& \tiny 0.24s & \tiny 0.70s & \tiny 90.73s & \tiny timeout\\  \hline \hline
&{ \tiny Non-conf points}  & \tiny conf & \tiny conf& \tiny 1 3 6 2 6 4 5 3 4 5 & \tiny p4 \\  \cline{2-6}
{\textbf CPUT4} & \tiny T(s)& \tiny 0.96s & \tiny 1.06s & \tiny  1.26s & \tiny 100.22s \\ \hline \hline
&{ \tiny Non-conf points}  & \tiny conf & \tiny---& \tiny 6 4 5 3 4 5 2 6 3 1 & \tiny --- \\  \cline{2-6}
{\textbf P} & \tiny T(s)& \tiny 3.01s & \tiny timeout & \tiny  0.17s & \tiny timeout \\ \hline \hline
\multicolumn{6}{|c|}{ \tiny p1 =  6 5 6 4 5 2 4 4 4 3 5 6 7 6 3 3 3 5 6 4 5 5 2 2 7 3 4 2 5 5 5 4 1 3 4 1 6 4 3 1 5 3 3 6 1 6 7 7 7 2 6 3 1 6 4 } \\
\multicolumn{6}{|c|}{ \tiny p2 =  7 1 6 3 4 6 1 7 3 2 5 1 7 3 5 4 2 6 6 6 4 3 6 5 3 4 4 2 4 6 1 3 7 5 5 2 5 5 3 7 6 3 1 6 4 3 5 4 2 4 6 5 5 4 3 } \\
\multicolumn{6}{|c|}{ \tiny p3 =  4 3 1 5 6 5 5 1 2 4 2 3 6 6 6 3 2 5 2 1 7 4 4 4 3 3 3 5 4 3 6 4 6 6 4 1 7 3 1 5 6 4 2 5 7 6 3 5 5 6 7 4 3 7 5   } \\
\multicolumn{6}{|c|}{ \tiny p4 =  1 3 6 2 5 4 3 5 2 6 4 5 3 4 5 2 6 3 5 4 4 5 3 7 6 4 1 3 6 7 1 7 6 3 1 4 6 7 5 2 6 3 1 7 6 4 5 4 3 5 4 6 2 5 3   } \\ \hline
\end{tabular}
\end{table}
Tab.\ref{CSeq} gives the results of CPTEST on two instances of the problem: an assembly line of 10 cars, 6 class and 5 options~; 
an assembly line with 55 cars, 7 class and 5 options.
Using $conf_{one}$, CPTEST reports non-conformities for the three first CPUT in less than 1sec for both instances. 
CPUT4 has no solution as the fault introduced on the {\tt pack} constraint prunes dramatically the search space. This case is
interesting as detecting this fault is really uneasy. With the $conf_{all}$ relation, the results are balanced as three instances
were not detected as non-conformant within the allocated time slot.
For example, in CPUT2, the capacity constraint of the first option is violated ({\tt 1 out of 2}). This fault results from a bad 
formulation but it is quickly detected with $conf_{one}$. When $conf_{all}$ is selected, more constraints have to be negated 
and then our algorithm has to backtrack a lot, which explains the failure. 
The non-conformity reached in this case satisfies the model-oracle and violates CPUT2, so it represents a correct assembly 
line that CPUT2 excludes from its solutions. Therefore, we can conclude that CPUT2 adds and removes solutions which make it difficult
to detect as non-conform.

\section{Conclusion}
\label{sec-conclusion}

In this paper, we introduced for the first time a testing framework that
is adapted to standard CP development processes. The
framework is built on solid notions such as conformity relations, oracles and test purposes
that are specific to CP. We also presented CPTEST an implementation of our framework dedicated
to the testing of OPL programs and evaluated it on difficult instances of two well-known
constraint problems, namely the Golomb ruler and car-sequencing problem. Our experimental
evaluation shows that CPTEST can efficiently detect non-trivial faults in faulty versions of
those two problems. 
A desirable extension of our framework and tool concerns its
application to other more open CP plateforms. In particular, we would like to apply our
conformity relations, oracles and testing notions to GECODE or CHOCO programs as we
could intervene on the core constraint solver of these systems. Developing notions of test
coverage similar of those that can be found in conventional programming requires instrumenting
the solver, something that was just not possible with the black-box solver of OPL.

\bibliographystyle{plain}
\bibliography{cp10}

\end{document}